\documentclass[journal,twoside]{IEEEtran}

\usepackage{graphicx}
\usepackage{graphicx,subfigure} 
\usepackage{cite} 
\usepackage{multirow} 
\usepackage{tabularx,booktabs} 
\usepackage{rotating}
\usepackage{threeparttable, tablefootnote}
\usepackage{color,soul}
\usepackage{comment}
\usepackage{amsmath}      
\usepackage{xcolor} 
\usepackage[colorlinks = true,
            linkcolor = black,
            urlcolor  = black,
            urlbordercolor=blue,
            citecolor = black,
            anchorcolor = black]{hyperref}
\usepackage[acronyms,nonumberlist,nopostdot,nomain,nogroupskip]{glossaries}
\usepackage{boldline}
\usepackage{hhline}
\newcommand{\virg}[1] {``{{#1}}''}
\newcommand{\red}[1] {{\color{black}{#1}}}

\newacronym{3gpp}{3GPP}{3rd Generation Partnership Project}
\newacronym{5g}{5G}{5th Generation}
\newacronym{5gacia}{5G-ACIA}{5G Alliance for Connected Industries and Automation}
\newacronym{6g}{6G}{6th Generation}
\newacronym{ADC}{ADC}{Analog-to-Digital Converter}
\newacronym{agv}{AGV}{Automated Guided Vehicle}
\newacronym{ai}{AI}{Artificial Intelligence}
\newacronym{ad}{AD}{Application Domain}
\newacronym{bs}{BS}{Base Station}
\newacronym{DAC}{DAC}{Digital-to-Analog Converter}
\newacronym{ed}{ED}{End Device}
\newacronym{gw}{GW}{Gateway}
\newacronym{ict}{ICT}{Information and Communications Technologies}
\newacronym{iiot}{IIoT}{Industrial Internet of Things}
\newacronym{isac}{ISAC}{Integrated Sensing and Communication}
\newacronym{iot}{IoT}{Internet of Things}
\newacronym{kpi}{KPI}{Key Performance Indicator}
\newacronym{llm}{LLM}{Large Language Model}
\newacronym{MADRL}{MADRL}{Multi-Agent Deep Reinforcement Learning}
\newacronym{ms}{MS}{Message Size}
\newacronym{MCP}{MCP}{Mobile Control Panel}
\newacronym{nr}{NR}{New Radio}
\newacronym{nlos}{NLoS}{Non-Line-of-Sight}
\newacronym{npn}{NPN}{Non-Public Network}
\newacronym{plc}{PLC}{Programmable Logic Controller}
\newacronym{ppp}{PPP}{Poission Point Process}
\newacronym{sbr}{SBR}{Service Bit Rate}
\newacronym{uav}{UAV}{Unmanned Aerial Vehicle}
\newacronym{ue}{UE}{User Equipment}
\newacronym{urllc}{URLLC}{Ultra-Reliable Low-Latency Communications}
\newacronym{st}{ST}{Survival Time}
\newacronym{scada}{SCADA}{Supervisory Control and Data Acquisition}
\newacronym{uc}{UC}{Use Case}
\newacronym{csi}{CSI}{Communication Service Interface}
\newacronym{pdc}{PDC}{Periodic Deterministic Communication}
\newacronym{adc}{ADC}{Aperiodic Deterministic Communication}
\newacronym{ndc}{NDC}{Non-Deterministic Communication}
\newacronym{ris}{RIS}{Reconfigurable Intelligent Surfaces}
\newacronym{ti}{TI}{Transfer Interval}
\newacronym{thz}{THz}{TeraHertz}
\newacronym{e2e}{E2E}{End-to-End}
\newacronym{e2el}{E2EL}{End-to-End Latency}
\newacronym{csa}{CSA}{Communication Service Availability}
\newacronym{csr}{CSR}{Communication Service Reliability}
\newacronym{v2x}{V2X}{Vehicle-to-Everything}
\newacronym{uwb}{UWB}{Ultra-Wide Band}
\newacronym{gnss}{GNSS}{Global Navigation Satellite System}
\newacronym{tsn}{TSN}{Time-Sensitive Networking}
\newacronym{poc}{PoC}{Proof-of-Concept}
\newacronym{amr}{AMR}{Autonomous Mobile Robot}
\newacronym{rf}{RF}{Radio Frequency}
\newacronym{ran}{RAN}{Radio Access Network}
\newacronym{cn}{CN}{Core Network}

\begin{document}
%
\title{Research Directions and Modeling Guidelines \\ for Industrial Internet of Things Applications}

\author{Giampaolo Cuozzo, Enrico Testi, Salvatore Riolo, Luciano Miuccio, \\ Gianluca Cena, Gianni Pasolini, Luca De Nardis, Daniela Panno, \\ Marco Chiani, Maria-Gabriella Di Benedetto, Enrico Buracchini, Roberto Verdone
\thanks{This work has been submitted to the IEEE for possible publication. Copyright may be transferred without notice, after which this version may no longer be accessible.}
\thanks{This work was partially supported by the European Union under the Italian National Recovery and Resilience Plan (NRRP) of NextGenerationEU, partnership on “Telecommunications of the Future” (PE00000001 - program “RESTART”). \\ 
Giampaolo Cuozzo is with WiLab, CNIT. \\
Enrico Testi, Gianni Pasolini, Marco Chiani, and Roberto Verdone are with DEI of the University of Bologna. \\
Salvatore Riolo, Luciano Miuccio, and Daniela Panno are with DIEEI of the University of Catania.\\
Gianluca Cena is with IEIIT, CNR.\\
Luca De Nardis, and Maria-Gabriella Di Benedetto are with DIET of University La Sapienza.\\
Enrico Buracchini is with Telecom Italia.
}}


%

\maketitle

\begin{abstract}
The Industrial Internet of Things (IIoT) paradigm has emerged as a transformative force, revolutionizing industrial processes by integrating advanced wireless technologies into traditional procedures to enhance their efficiency. The importance of this paradigm shift has produced a massive, yet heterogeneous, proliferation of scientific contributions. However, these works lack a standardized and cohesive characterization of the IIoT framework coming from different entities, like the 3rd Generation Partnership Project (3GPP) or the 5G Alliance for Connected Industries and Automation (5G-ACIA), resulting in divergent perspectives and potentially hindering interoperability. To bridge this gap, this article offers a characterization of (i) the main IIoT application domains, (ii) their respective requirements, (iii) the principal technological gaps existing in the current literature, and, most importantly, (iv) we propose a systematic approach for assessing and addressing the identified research challenges. Therefore, this article serves as a roadmap for future research endeavors, promoting a unified vision of the IIoT paradigm and fostering collaborative efforts to advance the field.
\end{abstract}

\IEEEpeerreviewmaketitle

\section{Introduction}

The advent of the fourth industrial revolution, commonly referred to as Industry 4.0, is revolutionizing the manufacturing landscape through the integration of \gls{ict} \cite{aceto2019survey}. This paradigm shift is characterized by the digitalization of production processes, and the adoption of cyber-physical systems that seamlessly blend physical assets with digital technologies. In this regard, the \gls{iiot} emerges as a pivotal facilitator within the Industry 4.0 framework, driving the Industry 4.0 paradigm from traditional, rigid industrial processes to interconnected, adaptive, and more efficient systems, by envisioning the adoption of wireless communications in traditional industrial processes to augment their efficiency and security \cite{5gaciaiiot}. \red{Industrial assets, such as robotic arms or valves, are equipped with wireless devices to support applications like analytics, monitoring, and control (detailed descriptions are provided in \cite{3gpp22804}).}


In light of this, factories pivot away from the traditional reliance on extensive wired communication technologies, embracing \red{wireless solutions that unlock unprecedented levels of flexibility and adaptability \cite{aceto2019survey}. Wi-Fi\textsuperscript{\textcopyright}, for instance, delivers high performance and user-friendly operation at a low cost but lacks energy efficiency, making it impractical for large-scale deployments involving battery-powered devices. LoRa\textsuperscript{\textcopyright}, on the other hand, offers long-range, low-power connectivity, making it well-suited for non-real-time monitoring applications, though its network throughput is severely limited. Similarly, ZigBee\textsuperscript{\textcopyright} and Bluetooth\textsuperscript{\textcopyright} are effective for short-range, low-power communication, ideal for specific applications like asset tracking or environmental monitoring, but they also struggle with limited data throughput \cite{aceto2019survey}. \gls{5g} \gls{nr} also emerged as a viable candidate for industrial communications thanks to key features such as \gls{urllc}, the \gls{npn} paradigm, and integration with the
widely adopted \gls{tsn} standard \cite{3gpp22104, 5gaciaiiot}; as of today, however, it remains both complex and expensive 
to deploy and maintain. Overall, the real-world adoption of these technologies remains limited because they are unable to comprehensively address all the diverse, stringent, and co-located requirements inherent in industrial environments.}

\red{This limitation highlights the necessity for further innovation and sets the stage for the development of next-generation mobile radio networks. In this regard, the academic world is already intensively shaping such evolution}, by proposing important paradigm shifts, like the use of the THz band, radar in next-generation \glspl{bs}, or \gls{ai}-radio communication interplay.


\begin{figure*}[!t]
\centering
	\includegraphics[width=\linewidth]{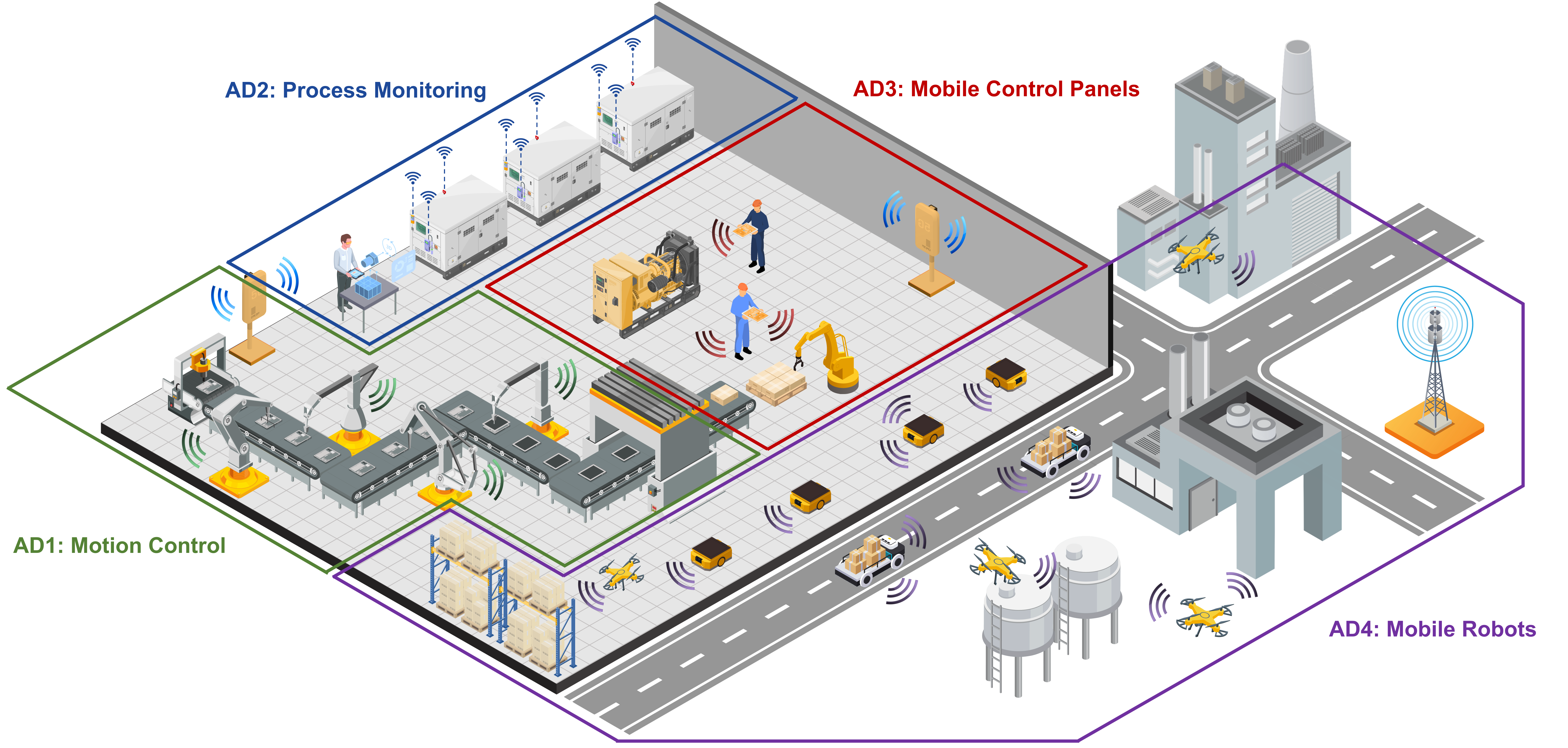}
	\caption{Illustration of the main IIoT ADs: in green, an example of motion control system implemented in an assembly line (AD1); in blue, a plethora of sensors gathering data from the production plant (AD2); in red, mobile control panels handled by workers and controlling industrial machinery (AD3); in purple, fleets of mobile robots interacting with the factory plant and extending network coverage to outdoor remote areas (AD4).}
	\label{fig:use-cases}
\end{figure*}

Despite this ambitious evolutionary trend, the majority of contributions to the literature do not account for the specificities of industrial environments (e.g., heterogeneous traffic), potentially limiting their practical value in that context. 

This may be attributed to the lack of a standardized and cohesive characterization of the \red{next-generation} \gls{iiot} framework, since different entities, such as \gls{3gpp} and \gls{5gacia}, propose their own perspective and nomenclature for \gls{iiot} application domains and \glspl{kpi}.

To address this ambiguity, this article aims to:

\begin{enumerate}
    \item Provide a \red{holistic description} of the main characteristics of industrial contexts, in terms of nomenclature, \glspl{ad}, \glspl{kpi}, and traffic types;
    \item Characterize the \gls{iiot} requirements for \red{future wireless} communications by performing a meticulous examination of the literature, encompassing white papers and standards, and by unifying their divergent aspects (e.g., consolidating different terminologies that refer to the same \gls{kpi});
    \item Identify the principal technological gaps in the current literature and present our perspective on the emerging research themes across various layers of the ISO-OSI standard;
    \item Propose guidelines for assessing and addressing the identified research challenges to promote reproducibility and compatibility among different scientific contributions.    
\end{enumerate}

\section{IIoT Application Domains}
\label{sec:iiot_application_domains}


\red{In this section, we clarify the different and sometimes contrasting nomenclatures used in the literature to describe application domains for the \gls{iiot}. Our goal, rather than proposing modifications or additions to existing naming conventions, is to provide a clearer understanding of these terms and to facilitate their consistent usage across various contexts.}

\subsection{Nomenclature}
One of the main characteristics of the \gls{iiot} literature is the presence of multiple synonyms, which may easily lead to misunderstanding. To solve this ambiguity, we hereby provide a list, in the form of a dictionary including the most common synonyms, that aims to merge the various \gls{iiot} terms and also constitutes the common nomenclature for this article:

\begin{itemize}
    \item \textit{Industrial Application/\gls{uc}:} It indicates a single, specific, and self-contained industrial application, e.g., assembly lines, thermal monitoring, etc.;
    \item \textit{Industrial \gls{ad}:} It indicates a family of industrial \glspl{uc}, such as motion control, process monitoring, etc.; 
    \item \textit{Factory/Manufacturing Plant/Production Plant/Industry Plant:} It indicates an industrial facility made of one or more buildings;
    \item \textit{Industrial Asset:} It refers to any element present within a factory or building (e.g., robotic arms, pumps, valves, pistons, etc.); 
    \item \textit{Device/\gls{ue}/Node/Tag:} It denotes a wireless-enabled component (e.g., a microcontroller board embedding sensors and/or actuators) associated with a given industrial asset of a \gls{uc}, such as an \gls{agv} transporting goods all over the factory, to monitor and/or control its operation;
    \item \textit{\gls{bs}:} It denotes a fixed radio transceiver that connects wireless devices, whether fixed or mobile, in its coverage area to the network infrastructure, serving as a pivotal node for signal transmission and reception.
\end{itemize}

\subsection{AD1: Motion control}
Motion control is among the most challenging and demanding closed-loop control application domains in industrial environments\cite{3gpp22804}. A motion control system (see Fig.~\ref{fig:use-cases}) is responsible for controlling the movement of industrial assets in a well-defined manner and is composed of three main elements: sensor(s), actuator(s), and a controller. The sensors gather data from the environment and send the actual values to the motion controller. In the opposite direction, the motion controller sends a command to the actuators which thereupon perform the corresponding action(s) on one or several processes (this may also include simple software updates).

Different motion controllers may also need to communicate to cooperate in a coordinated and tightly coupled manner when performing a shared task \cite{3gpp22804}.

\textit{Example of \glspl{uc}}: Machine tools, packaging machines, printing machines, assembly lines.

\subsection{AD2: Process Monitoring} 
Process monitoring is an essential aspect of any industrial plant, allowing the tracking of operations and assets to enhance productivity and product quality, while concurrently minimizing waste, downtime, and energy consumption.

As depicted in Fig.~\ref{fig:use-cases}, data concerning production processes is gathered through a myriad of devices strategically placed in the plant and functions such as the well-known \gls{scada}. This data is then transmitted through the network, often at regular intervals, to designated data repositories (located either on local premises or in the cloud). Subsequently, the information becomes available for processing tasks, such as data visualization and predictive maintenance, empowering industry owners with valuable insights.

\textit{Example of \glspl{uc}}: Temperature, vibration, or thermal monitoring.

\subsection{AD3: Mobile Control Panels} 
Industrial assets currently rely on control panels known as \glspl{MCP}. These devices are intended for configuring, monitoring, debugging, controlling, and maintaining machines, robots, cranes, or production lines (see Fig.~\ref{fig:use-cases}). \glspl{MCP} might also incorporate safety features, such as an emergency button to halt machine operation in hazardous situations, preventing harm to individuals or damage to assets. For the same reason,
\glspl{MCP} may need to support the special \virg{enabling device mode} as well. In this scenario, the operator must manually keep the enabling device switch in a specific stationary position. Failure to do so will result in the controlled equipment promptly moving to a safe stationary position, guaranteeing that
the operator’s hand(s) stay on the panel (rather than under a molding press, for example), and that the operator does not suffer from any electric shock or the like.


\textit{Example of \glspl{uc}}: Remote control, or emergency stop.

\subsection{AD4: Mobile Robots} 

Mobile robots will play a crucial role in next-generation industrial environments thanks to their capacity to assist in work processes, transport goods and assets, extend radio coverage, and provide support and rescue in safety-critical conditions and hard-to-reach areas. Mobile robots are typically divided into \glspl{agv}, ground units primarily used in indoor environments, and \glspl{uav} that can be deployed both indoors and outdoors, as depicted in Fig.~\ref{fig:use-cases}. Usually, one or more centralized units (e.g., the guidance control system) facilitate bidirectional traffic exchange with the robots, overseeing navigation and data collection. Direct inter-robot communications might be necessary for cooperative task completion. Additionally, robots may need to interact with (or control) machinery inside or outside the buildings to enhance production and ensure security standards.

\textit{Example of \glspl{uc}}: Video-operated remote control, video streaming, cooperative driving.

\section{IIoT Characterization}
\label{sec:kpi}

\begin{table*}[!t]
\caption{List of communication service and positioning KPIs \textcolor{black}{\cite{5gaciaiiot,3gpp22261,3gpp22104,3gpp22872}}.}
\label{table_KPIs}
\begin{tabular}{c|c|c|c|c|}
  \cline{2-5}

   &
  \begin{tabular}[c]{@{}c@{}} \multirow{2}{*}{\textbf{KPI name}}    \end{tabular} 
   &
  \begin{tabular}[c]{@{}c@{}} \multirow{2}{*}{\textbf{KPI acronym}}    \end{tabular}  
   &
    \begin{tabular}[c]{@{}c@{}} \multirow{2}{*}{\textbf{KPI description}}    \end{tabular}  
   &
    \begin{tabular}[c]{@{}c@{}} \multirow{2}{*}{\textbf{Unit}}    \end{tabular}  \\ 
  &&&&\\
  \cline{1-5}
  
   \multicolumn{1}{|l|}{}
   &
   \multirow{4}{*}{\begin{tabular}[c]{@{}c@{}}  End-to-End latency $/$ \\ Latency\end{tabular}}
   &
  \multirow{4}{*}{\begin{tabular}[c]{@{}c@{}} E2EL \end{tabular}}
  &
   \multirow{4}{*}{\begin{tabular}[c]{@{}c@{}} \textcolor{black}{Time allotted to transfer a message from the moment it is transmitted by the source}\\ \textcolor{black}{CSI to the moment it is successfully received at the destination CSI.} \\ \textcolor{black}{It is a one-way latency, thus the loop-back latency
   would be twice the E2EL}
   \end{tabular}} &
   \multirow{4}{*}{ms}  \\
    \multicolumn{1}{|l|}{}&&&&\\
  \multicolumn{1}{|l|}{}&&&&\\
    \multicolumn{1}{|l|}{}&&&&\\
 \cline{2-5}
  



\multicolumn{1}{|l|}{}&
   \multirow{4}{*}{\begin{tabular}[c]{@{}c@{}}  Message Size $/$  \\ Typical Payload Size \end{tabular}}   &
    \multirow{4}{*}{MS} &
    \multirow{4}{*}{\begin{tabular}[c]{@{}c@{}}The maximum size of the user data packet delivered \\ from the application to the ingress of the communication system and from \\ the egress of the communication system to the application\end{tabular}}
  &
    \multirow{4}{*}{Bytes} \\ 
    \multicolumn{1}{|l|}{}&&&&\\
    \multicolumn{1}{|l|}{}&&&&\\
    \multicolumn{1}{|l|}{}&&&&\\
  \cline{2-5}
  
\multicolumn{1}{|l|}{}& \multirow{3}{*}{\begin{tabular}[c]{@{}c@{}} Transfer Interval $/$  \\ Cycle Time \end{tabular}} &
\multirow{3}{*}{TI} 
&
\multirow{3}{*}{\begin{tabular}[c]{@{}c@{}}Time difference between two consecutive transfers of data 
\\ from an application to the wireless communication system
via the service interface \end{tabular}}
&
\multirow{3}{*}{ms} \\


    \multicolumn{1}{|l|}{}&&&&\\
    \multicolumn{1}{|l|}{}&&&&\\
    \cline{2-5}

  
  
\multicolumn{1}{|l|}{\multirow{-5}{*}{\begin{turn}{90} Communication service \end{turn}}} & \multirow{4}{*}{Survival Time} & \multirow{4}{*}{ST} & \multirow{4}{*}{\begin{tabular}[c]{@{}c@{}}Maximum time interval over which the communication service can fail to meet the \\ application message delay requirement before an application layer failure \\occurs, leading to the communication service being considered as unavailable \end{tabular}} &
   \multirow{4}{*}{ms} \\ 
   \multicolumn{1}{|l|}{}&&&&\\
   \multicolumn{1}{|l|}{}&&&&\\
   \multicolumn{1}{|l|}{}&&&&\\
  \cline{2-5}

\multicolumn{1}{|l|}{}& \multirow{4}{*}{\begin{tabular}[c]{@{}c@{}} Communication Service \\ Availability $/$  \\ Availability\end{tabular}} &
  \multirow{4}{*}{CSA} &   \multirow{4}{*}{\begin{tabular}[c]{@{}c@{}}Amount of time during which the E2E communication service is \\ delivered according to the agreed-upon QoS, divided by the total time \\ in which the system is expected to deliver the E2E service\end{tabular}} & \\ 
     \multicolumn{1}{|l|}{}&&&&\\
   \multicolumn{1}{|l|}{}&&&&\\
   \multicolumn{1}{|l|}{}&&&&\\
   \cline{2-5}
\multicolumn{1}{|l|}{}  
&\multirow{4}{*}{\begin{tabular}[c]{@{}c@{}}Communication Service \\ Reliability $/$  \\ Reliabity\end{tabular}} &
  \multirow{4}{*}{CSR} &   \multirow{4}{*}{\begin{tabular}[c]{@{}c@{}}Mean value of time intervals between consecutive\\unavailability events for the communication service \end{tabular}} &
\multirow{4}{*}{year}   \\ 
     \multicolumn{1}{|l|}{}&&&&\\
   \multicolumn{1}{|l|}{}&&&&\\
   \multicolumn{1}{|l|}{}&&&&\\
  
  \cline{2-5}
  
\multicolumn{1}{|l|}{}&
\multirow{3}{*}{\begin{tabular}[c]{@{}c@{}}Service Bit Rate $/$  \\ User Experienced Data Rate\end{tabular}}&
  \multirow{3}{*}{SBR} &
  \multirow{3}{*}{\begin{tabular}[c]{@{}c@{}} Minimum rate measured at the CSI required to achieve a \\ satisfactory user experience \end{tabular}} 
  &
  \multirow{3}{*}{Mbps} \\ 
     \multicolumn{1}{|l|}{}&&&&\\
   \multicolumn{1}{|l|}{}&&&&\\
\hhline{|=====|}
    
\multicolumn{1}{|l|}{}   & \multirow{3}{*}{Horizontal Accuracy} &  \multirow{3}{*}{HA} &   \multirow{3}{*}{\begin{tabular}[c]{@{}c@{}}Distance between the estimated position and the actual position \\ (ground truth) in the horizontal x-y plane\end{tabular}} &  \multirow{3}{*}{m} \\ 
     \multicolumn{1}{|l|}{}&&&&\\
   \multicolumn{1}{|l|}{}&&&&\\
  \cline{2-5}

\multicolumn{1}{|l|}{\multirow{1}{*}{\begin{turn}{90} Positioning service\end{turn}}}   &\multirow{3}{*}{Vertical Accuracy} & \multirow{3}{*}{ VA} &  \multirow{3}{*}{\begin{tabular}[c]{@{}c@{}}Distance between the estimated position and the actual position \\ (ground truth) in the vertical axis z\end{tabular}} & \multirow{3}{*}{ m} \\
     \multicolumn{1}{|l|}{}&&&&\\
   \multicolumn{1}{|l|}{}&&&&\\
  \cline{2-5}
  
\multicolumn{1}{|l|}{}   & \multirow{3}{*}{\begin{tabular}[c]{@{}c@{}}Heading Accuracy  $/$  \\ Heading \end{tabular}}  &   \multirow{3}{*}{HEA} &   \multirow{3}{*}{\begin{tabular}[c]{@{}c@{}}Absolute value of difference between the measured direction of movement \\and actual direction\end{tabular}} &
  \multirow{3}{*}{degrees} \\ 
       \multicolumn{1}{|l|}{}&&&&\\
   \multicolumn{1}{|l|}{}&&&&\\
  \cline{2-5}

\multicolumn{1}{|l|}{}   &\multirow{2}{*}{Speed Accuracy} &  \multirow{2}{*}{SA} &  \multirow{2}{*}{\begin{tabular}[c]{@{}c@{}}Absolute value of the difference between measured speed and actual speed\\ \end{tabular}} &
  \multirow{2}{*}{m/s} \\ 
     \multicolumn{1}{|l|}{}&&&&\\
   \cline{2-5}
  
\multicolumn{1}{|l|}{}   &\multirow{4}{*}{\begin{tabular}[c]{@{}c@{}}Positioning Service \\ Availability $/$  \\ Availability\end{tabular}} &
  \multirow{4}{*}{PSA} &  \multirow{4}{*}{\begin{tabular}[c]{@{}c@{}}Percentage of time in which the positioning service can satisfy  a positioning \\request\end{tabular}} &  \multirow{4}{*}{\%} \\ 
         \multicolumn{1}{|l|}{}&&&&\\
      \multicolumn{1}{|l|}{}&&&&\\
      \multicolumn{1}{|l|}{}&&&&\\
    \cline{2-5}
  
\multicolumn{1}{|l|}{}   &\multirow{4}{*}{\begin{tabular}[c]{@{}c@{}}Positioning Service\\  Latency $/$  \\ Latency\end{tabular}} &
  \multirow{4}{*}{PSL} &  \multirow{4}{*}{\begin{tabular}[c]{@{}c@{}}Time between a positioning request and the corresponding reply by the \\positioning service\end{tabular}} &  \multirow{4}{*}{ms} \\
           \multicolumn{1}{|l|}{}&&&&\\
      \multicolumn{1}{|l|}{}&&&&\\
      \multicolumn{1}{|l|}{}&&&&\\
  
  \cline{1-5}
\end{tabular}
\end{table*}

\textcolor{black}{We then hereby describe the main \glspl{kpi}, the typical traffic types, and requirements to be adopted in the \gls{iiot} paradigm.}

\subsection{KPIs}
The \glspl{uc} of the \glspl{ad} described in the previous section cover diverse requirements, which has prompted both \gls{3gpp} and \gls{5gacia} to provide various \glspl{kpi} instrumental for \gls{iiot} applications. These \glspl{kpi} encompass several metrics essential for assessing the performance of industrial communication systems. 
To describe them, we employ metrics specified in  \gls{3gpp} documents\cite{3gpp22261,3gpp22104,3gpp22872}, and by \gls{5gacia} in its documents (e.g., \cite{5gaciaiiot}). 
A weakness identified in the collective analysis of the aforementioned documents is the use of different terminologies to refer to the same concept. To consistently address the definition of \glspl{kpi}, we assemble the nomenclature used by both entities. The resulting comprehensive list of \glspl{kpi} is detailed in Table \ref{table_KPIs}, and incorporates names adopted in \cite{3gpp22261,3gpp22104,3gpp22872,5gaciaiiot}, acronyms used in this article, along with their respective definitions and units.
These \glspl{kpi} have been grouped into two macro-categories related to the communication service and the positioning service.

The former category is related to the \gls{csi}, i.e., the interface between the industrial application and the underlying wireless communication system. This interface is the primary reference point where different traffic flows, their principal requirements, and associated \glspl{kpi} are defined. Notably, there is an important difference in the metric definition between the industrial application (a.k.a., communication service) and the wireless communication system.
Specifically, referring to the wireless communication system, \virg{reliability} is measured as the percentage of successfully delivered packets within the required time frame, focusing on individual packet losses without considering their temporal distribution. However, industrial applications often prioritize consecutive packet losses differently.
Losing a single packet may minimally impact the network performance, while losing several consecutive packets could lead to application failure. To address this, specific \glspl{kpi} are introduced in the communication service perspective (see Table \ref{table_KPIs}). The \gls{st} is introduced to quantify the time an application service can tolerate without receiving an expected message, allowing to define \gls{csa} and \gls{csr}. 
It is worth noting that we intentionally excluded the jitter as it is typically expressed as a function of other \glspl{kpi} (e.g., \gls{e2el}) \cite{3gpp22804, 3gpp22104, 3gpp22261}.

Regarding positioning, it is typically provided as a dedicated service. It may run either on top of the communication infrastructure or on a dedicated positioning infrastructure, such as a \gls{gnss} or an \gls{uwb} positioning system. The dedicated set of \glspl{kpi} describes the requirements imposed on the positioning service in terms of availability and accuracy of the position information and motion (i.e., speed and heading).

\subsection{Traffic types}

The \glspl{kpi} listed in Table \ref{table_KPIs} refer to diverse traffic flows at the \gls{csi}. Specifically, traffic flows can be categorized into three types, described as follows. 
\begin{itemize}
 \item\textbf{\gls{pdc}} refers to sending messages of size \gls{ms} at a constant \gls{ti} with stringent requirements on both  \gls{e2el} and \gls{csa}.
\item\textbf{\gls{adc}} consists of messages sent aperiodically, i.e., there is no fixed \gls{ti}  between consecutive messages. Despite the lack of a pre-set sending time, the requirements for \gls{e2el} and \gls{csa}  remain stringent \cite{3gpp22104}. Note that, for this specific traffic type, \gls{ms}, \gls{ti}, and \gls{st} are not defined. Therefore, its characterization relies solely on the \gls{sbr}.

\item\textbf{\gls{ndc}} encompasses all traffic types other than periodic/aperiodic deterministic communication. In this case, there are no requirements either in terms of \gls{e2el} or \gls{csa}. As in the case of \gls{adc}, the traffic is defined through \gls{sbr}.
\end {itemize}

\begin{table*}[]
\renewcommand{\arraystretch}{0.9}
\caption{CSI requirements for different ADs \textcolor{black}{\cite{3gpp22261,3gpp22104}}.}
\centering

\label{table_requirements}
\begin{tabular}{|c|c|c|c|c|c|c|c|c|c|}

\hline
&&&&&&&&& \\
\textbf{\begin{tabular}[c]{@{}c@{}}Application \\ Domain (AD) \end{tabular}} 
& \textbf{\begin{tabular}[c]{@{}c@{}}Use Case\\ (UC)\end{tabular}} 
& \textbf{\begin{tabular}[c]{@{}c@{}}Traffic \\ type\end{tabular}} 

& \textbf{\begin{tabular}[c]{@{}c@{}} E2EL\\ $[\text{ms}]$ \end{tabular}} 

& \textbf{\begin{tabular}[c]{@{}c@{}}MS \\  $[\text{Bytes}]$ \end{tabular}}

& \textbf{\begin{tabular}[c]{@{}c@{}}TI \\  $[\text{ms}]$ \end{tabular}} 

& \textbf{\begin{tabular}[c]{@{}c@{}}ST \\  $[\text{ms}]$ \end{tabular}} 

&\textbf{\begin{tabular}[c]{@{}c@{}} CSA \\   \end{tabular}}  
& \textbf{\begin{tabular}[c]{@{}c@{}} CSR \\  $[\text{years}]$ \end{tabular}}   
& 

\textbf{\begin{tabular}[c]{@{}c@{}}SBR \\  $[\text{Mbps}]$ 
\end{tabular}} \\
 &&&&&&&&&\\
\hline
 &&&&&&&&&\\
 & \textit{Machine tool} &  & $<$ TI & 50 & 0.5 & 0.5 & $1-[10^{-5} , 10^{-7}]$ &$\approx 10$ & N/A \\ 
 &&&&&&&&&\\
 
 & \textit{\begin{tabular}[c]{@{}c@{}}Packaging\\ machine\end{tabular}} &  & $<$ TI & 40 & 1 & 1 &$1-[10^{-6} , 10^{-8}]$ & $\approx$ 10 & N/A \\ 
 &&PDC&&&&&&&\\
 \textit{\begin{tabular}[c]{@{}c@{}}AD1: Motion\\ Control\end{tabular}}& \textit{Printing machine} &  & $<$ TI & 20 & 2 & 2 & $1-[10^{-6} , 10^{-8}]$ & $\approx$ 10 & N/A \\
 &&&&&&&&&\\
 & \textit{\begin{tabular}[c]{@{}c@{}}Machine in\\ assembly line\end{tabular}} &  & $<$ TI & 1000 & $\leq$ 50 & 50 & $1-[10^{-6} , 10^{-8}]$ & $\approx 10$ & N/A \\ 
  &&&&&&&&&\\
\cline{3-10}
 &&&&&&&&&\\
 & \textit{\begin{tabular}[c]{@{}c@{}}Software\\ updates\end{tabular}} & NDC & N/A & N/A & N/A & N/A & N/A & $\approx \frac{1}{12}$ &  $\geq 1$ \\
  &&&&&&&&&\\
 \hline
  &&&&&&&&&\\
\textit{\begin{tabular}[c]{@{}c@{}}\end{tabular}} & \textit{\begin{tabular}[c]{@{}c@{}}Temperature\\ sensor\end{tabular}} &  & $<$ 100 & 20 & \begin{tabular}[c]{@{}c@{}}100 to\\ 60000\end{tabular} & 3 $\times$ TI & $1 - 10^{-4}$ & $\geq \frac{1}{54}$ & N/A \\ 
 &&&&&&&&&\\
\textit{\begin{tabular}[c]{@{}c@{}}AD2: Process\\ Monitoring\end{tabular}} & \textit{\begin{tabular}[c]{@{}c@{}}Vibration\\ sensor\end{tabular}} & PDC & $<$ 100 & 25000 & \begin{tabular}[c]{@{}c@{}}$\leq$ 1000\end{tabular} & 3 $\times$ TI & $1 - 10^{-4}$ & $\geq \frac{1}{54}$ & N/A \\ 
&&&&&&&&&\\
 \textit{\begin{tabular}[c]{@{}c@{}}\end{tabular}} & \textit{\begin{tabular}[c]{@{}c@{}}Thermal\\ camera\end{tabular}} &  & $<$ 100& 250000 & \begin{tabular}[c]{@{}c@{}}$\leq$ 1000\end{tabular} & 3 $\times$ TI & $1 - 10^{-4}$ & $\geq \frac{1}{54}$ & N/A \\ 

&&&&&&&&&\\
\hline
&&&&&&&&&\\
 & \textit{\begin{tabular}[c]{@{}c@{}}Remote control\\ of assembly\\ robots or milling\\ machines\end{tabular}} &  & $<$ TI & 40 to 250 & 4 to 8 & = TI & $1-[10^{-6} , 10^{-8}]$ & $\approx \frac{1}{12}$ & N/A \\
 &&&&&&&&&\\
 
 & \textit{\begin{tabular}[c]{@{}c@{}}Remote control\\ of mobile\\ cranes or mobile\\ pump\end{tabular}} & PDC & $<$ TI & 40 to 250 & $<$ 12 & 12 & $1 - 10^{-8}$ & $\approx 1$ & N/A \\ 
  \textit{\begin{tabular}[c]{@{}c@{}}AD3: Mobile control\\ panels\end{tabular}} &&&&&&&&&\\
 & \textit{\begin{tabular}[c]{@{}c@{}}Emergency stop\\ \end{tabular}} &  & $<$ 8 & 40 to 250 & 8 & 16 &$1-[10^{-6} , 10^{-8}]$ & $\approx \frac{1}{365}$ & N/A \\
  &&&&&&&&&\\
\cline{3-10}
 &&&&&&&&&\\
 & \textit{\begin{tabular}[c]{@{}c@{}}Data\\ transmission in \\ parallel to remote\\ control\end{tabular}} & ADC & $<$ 30 & N/A & N/A & N/A & $1-[10^{-6} , 10^{-8}]$ & $\approx \frac{1}{12}$ & $>$ 5 \\ 
  &&&&&&&&&\\
\hline
 &&&&&&&&&\\
 & \textit{\begin{tabular}[c]{@{}c@{}}Video-operated\\ remote control\end{tabular}} &  & $<$ TI & \begin{tabular}[c]{@{}c@{}}15000 to\\ 25000\end{tabular} & \begin{tabular}[c]{@{}c@{}}10 to\\ 100\end{tabular} & = TI &$< 1-10^{-6}$ & $\approx 1$ & N/A \\ 
  &&&&&&&&&\\
 & \textit{\begin{tabular}[c]{@{}c@{}}Cooperative\\ driving\end{tabular}} & PDC & $<$ TI & 40 to 250 & \begin{tabular}[c]{@{}c@{}}10 to\\ 50\end{tabular} & = TI &$< 1-10^{-6}$ & $\approx 10$ & N/A \\ 
 &&&&&&&&&\\
\textit{AD4: Mobile robots} & \textit{\begin{tabular}[c]{@{}c@{}}Machine \\ control\end{tabular}}&  & $<$ TI & 40 to 250 & 1 to 10 & = TI & $< 1-10^{-6}$ & $\approx 10$ & N/A \\ 
  &&&&&&&&&\\
 \cline{3-10}
  &&&&&&&&&\\
 & \textit{Video streaming} & ADC & $<$ 10 & N/A & N/A & N/A & $< 1-10^{-6}$&$\approx \frac{1}{54}$ & $>$ 10 \\
  &&&&&&&&&\\
 \cline{3-10}
  &&&&&&&&&\\
\multicolumn{1}{|l|}{} & \textit{\begin{tabular}[c]{@{}c@{}}Real-time video\\ streaming\\ to the guidance \\ control system\end{tabular}} & NDC & N/A & N/A & N/A & N/A & N/A & $\approx \frac{1}{12}$ & $>$  10 \\ 
 &&&&&&&&&\\
 \hline
\end{tabular}
\end{table*}

\begin{table*}
    \renewcommand{\arraystretch}{2.0}
    \caption{Positioning requirements for AD3 and AD4 \textcolor{black}{\cite{3gpp22872}}.}
    \label{positioning_requirements}
    \centering
        \begin{tabular}{|c|c|c|c|c|c|c|c|}
            \hline
            \textbf{Application Domain (AD)} & \textbf{Use Case (UC)} &  \textbf{HA $[\text{m}]$ } & \textbf{ VA $[\text{m}]$} &  \textbf{HEA} &  \textbf{SA} &  \textbf{PSA $[\text{\%}]$} &  \textbf{PSL $[\text{ms}]$} \\
            \hline
            \textit{AD3: Mobile control panels}&\textit{Multiple}& 1-5 &3 &10 & N/A&  90-99.9 & 1000-5000\\
            \hline
            \multirow{2}{*}{\textit{AD4: Mobile robots}} &\multicolumn{1}{c|}{\textit{Involving AGVs}}& 0.3 &3 &10 & 0.5&  99.9 & 10\\
            \cline{2-8}
             &\multicolumn{1}{c|}{\textit{Involving UAVs}} & 0.1 & 0.1 & 2 & 0.5 & 99.9 & 10 \\
            \hline
        \end{tabular}%
\end{table*}

\subsection{Requirements} 



The requirements at the \gls{csi} for the various \gls{iiot} \glspl{ad} and corresponding \glspl{uc} are summarized in Table \ref{table_requirements}. The term “N/A” therein represents “not applicable” and is used for \glspl{kpi} that are not defined.


Additionally, Table \ref{positioning_requirements} shows the positioning requirements for \gls{ad}3 and \gls{ad}4, in accordance with the \gls{3gpp} technical specifications that do not include \gls{ad}1 and \gls{ad}2. Specifically, for \gls{ad}3, a range of values is specified across various scenarios, while \gls{ad}4 requirements are categorized based on the type of industrial assets under consideration (\glspl{agv} vs \glspl{uav}), reflecting their distinct mobility characteristics. As can be observed from Table \ref{positioning_requirements}, \gls{ad}4 imposes notably more stringent requirements than \gls{ad}3.



\section{Research Directions and Modelling Guidelines}
\label{sec:research_directions_and_modelling_guidelines}
\textcolor{black}{In this section, we identify the key technological gaps in the current literature and offer our perspective on how emerging research fields could address these gaps to fulfill the \red{next-generation} \gls{iiot} requirements.}
\subsection{Research Directions}


After a detailed analysis of the \gls{iiot} \glspl{ad} and their characterization, we provide an overview of the major technology gaps in this field, focusing on the most promising research directions and emphasizing the pivotal challenges that will be addressed in the upcoming years. 

\noindent\textbf{Terahertz and Optical Communications.} In the last decades, \gls{thz} and optical communications have emerged thanks to their capability of achieving exceptionally high data rates in short-range scenarios, offering huge bandwidths and enabling high-throughput applications. These technologies can largely benefit industrial indoor communication scenarios as \gls{ad}1 and \gls{ad}3, which demand low \gls{e2el} and stringent \gls{csa}, while also catering to the high throughput needs of \gls{ad}2.

\textit{Open challenges:} Both \gls{thz} and optical communication systems face various challenges due to environmental factors and technical limitations \cite{SonLee:J22}.
First of all, such systems often rely on highly focused, narrow beams that must be accurately aligned to avoid signal losses. 
Overcoming this challenge requires innovative solutions in beam focusing, alignment mechanisms, and possibly adaptive systems that can adjust for misalignments in real-time. 
Moreover, molecular absorption in the \gls{thz} spectrum by water vapor or other molecules commonly found in industrial indoor environments can generate additional noise, affecting the quality and reliability of communication systems.

\noindent\textbf{Network as a Sensor}. \gls{thz} communications allow for the development of compact, miniaturized components that enable the integration of multiple antennas and transceivers in small devices (e.g., wearables), facilitating precise and distributed sensor deployment and data collection mechanisms.
In this context, next-generation industrial network paradigms envision leveraging the communication network as a sensor, to gather real-time data on industrial processes, environmental conditions, and machinery statuses. 
Moreover, it is expected that next-generation devices will possess \gls{isac} capabilities, leveraging on the high spatial resolution of mmWave and \gls{thz} technologies. The integration of communication and sensing paradigms, along with the capability to gather data from numerous distributed sensors form the basis for \gls{ad}2 and \gls{ad}3.
Furthermore, the high-precision positioning and tracking systems for vehicles and devices in both indoor and outdoor environments (e.g., \gls{ad}4) are essential to enable next-generation networks to meet the requirements and ensure safety standards.

\textit{Open challenges:} The high density of obstacles (e.g., machinery) in indoor industrial environments can hinder signal propagation causing reflections, scattering, and attenuation. The development of ad-hoc network layout design paradigms and innovative signal processing techniques is crucial for guaranteeing high communication and sensing performance. Recently, \gls{ris} have emerged as a means to optimize communication links and improve sensing coverage, nevertheless introducing an additional layer of complexity to network management \cite{KisChaOtt:J21}. 
Moreover, localizing and tracking objects transitioning from outdoor to indoor scenarios (and vice versa), as well as combining multi-technologies indoor localization techniques to meet the desired requirements, are still open problems.

\noindent\textbf{Massive Multiple Access}. The simultaneous connections of thousands of \gls{iot} devices can take advantage of small-cell or cell-free communication systems, in which cutting-edge low-power radio access techniques allow for higher coverage and capacity. While traditional macro cells might face limitations mainly due to propagation issues, small-cell/cell-free systems enable localized connectivity, providing reliable wireless communication in specific zones within the factory \cite{MahLopMoe:J21}. Such systems can help industrial networks meet the scalability and flexibility requirements even in dynamic environments with varying device densities (e.g., \gls{ad}2).

\textit{Open challenges:} Since the multitude of potentially interfering \gls{iot} devices usually send sporadic data to the \gls{bs} developing innovative protocols for massive random access is a promising solution to reduce network access time. However, among their primary unresolved issues there is the problem of channel estimation, which worsens when dealing with short packets subject to significant interference, and the substantial latency introduced by interference cancellation mechanisms.

\noindent\textbf{Artificial Intelligence}. The pervasiveness of \gls{ai} in contemporary industrial production plants (encompassing \gls{ad}1 to \gls{ad}4) is revolutionizing the way industries operate and optimize their processes \cite{PerJiaLee:J20}. Modern edge-\gls{ai} allows for data processing closer to the source, supporting quicker decision-making and reducing latency in time-sensitive applications.
Additionally, online learning algorithms enable systems to adapt and learn from dynamic environments, improving performance and resilience in changing conditions.
Furthermore, the revolution of \glspl{llm} can assist intent-based orchestration by interpreting and processing natural language commands or intents from human operators. This simplifies network management and configuration, enhancing operational efficiency.

\textit{Open challenges:} At the moment, the development of faster \gls{ai} models that can easily adapt to new dynamic scenarios without the need for extensive and energy-consuming re-trainings is one of the most important open challenges.

\noindent \textbf{Vehicular Communications}. Mobile robots,  particularly relevant to \gls {ad}4, will play a significant role in modern industries by supporting production processes. 
In particular, \glspl{uav} can act as mobile base stations or relays, establishing network connectivity even in areas with challenging terrain or temporary infrastructure needs, or during emergencies. 
To enable mobile robot operations, a robust backbone network ensuring efficient and dependable communication within the robots and the control systems is required. In this context, \gls{v2x} emerging standards facilitate communication among mobile robots, infrastructure, and other devices, improving safety and traffic management. In particular, the possibility to have sidelinks that allow devices in spatial proximity to communicate efficiently, forming ad-hoc networks in environments with limited infrastructure or in scenarios requiring rapid data exchange, is of paramount importance.

\textit{Open challenges:} Jointly optimizing the trajectories of mobile robots and managing radio resources while fulfilling application requirements and ensuring security standards is still an open problem. \gls{MADRL}-based solutions have been recently proposed; however, these techniques require long training phases and are not easily adaptable to new scenarios. Moreover, integrating \gls{v2x} standards into industrial wireless networks is one of the biggest challenges at the moment.



\begin{figure*}[!t]
\centering
	\includegraphics[width=\linewidth]{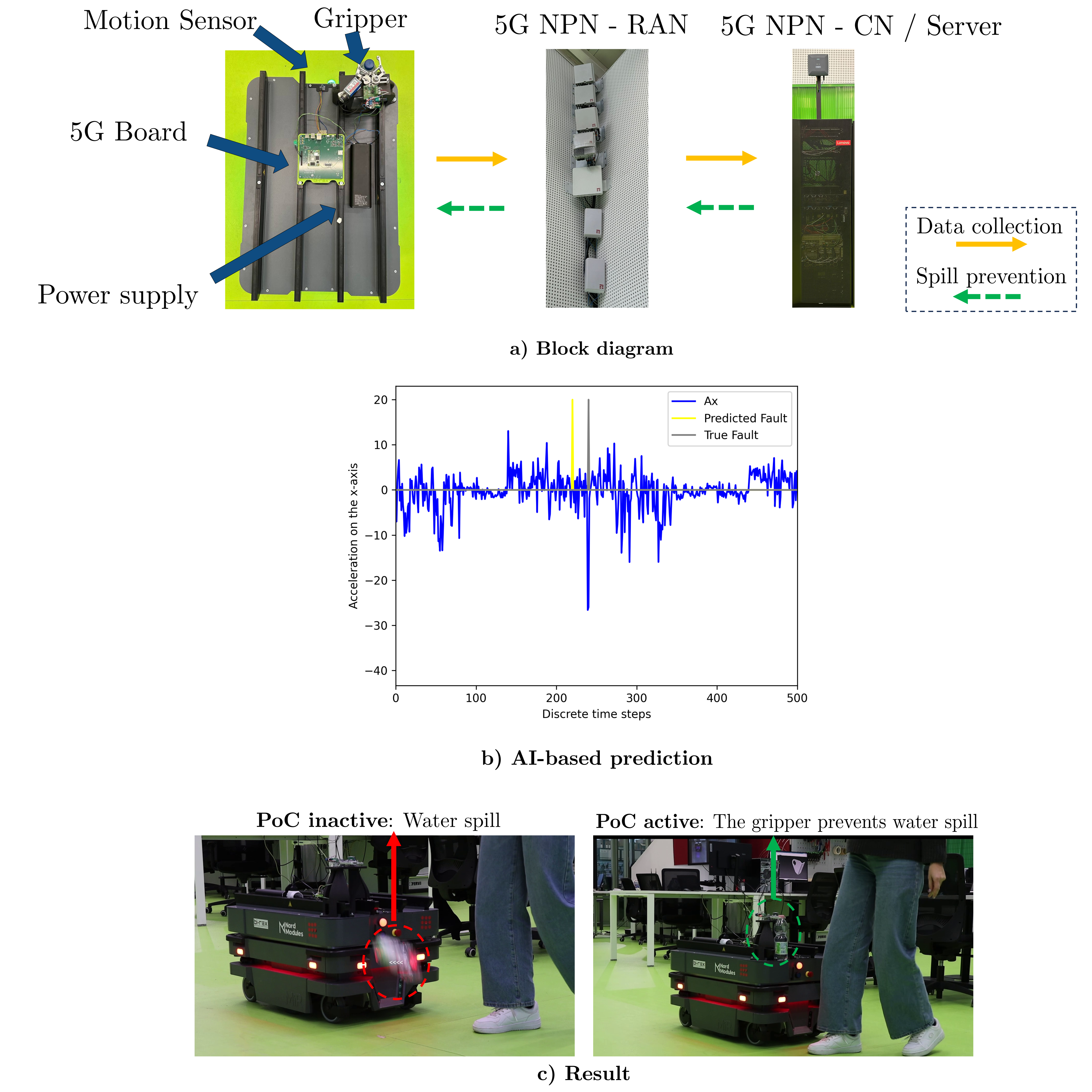}
	\caption{\red{Experimental \gls{poc} for \gls{ad}4. The focus of this \gls{poc} is on mitigating risks associated with an Autonomous Mobile Robot (\gls{amr}), which autonomously navigates industrial spaces and transports hazardous liquids. Although the \gls{amr} is equipped with emergency systems (e.g., proximity sensors and lasers) that trigger abrupt stops to avoid collisions, sudden braking can lead to liquid spills, posing a threat to human operators. The block diagram of the considered solution is illustrated in a). The \gls{amr} is equipped with a motion sensor, and the collected data are sent to a local \gls{5g} board powered by a power bank. The board transmits the data to a local server via a private \gls{5g} \gls{npn}, consisting of a dedicated Radio Access Network (\gls{ran}) and Core Network (\gls{cn}). The local server, housed within the same rack as the \gls{5g} \gls{cn}, hosts \gls{ai} algorithms that process real-time motion data to predict potential spills. An example of a prediction is shown in b), where the x-axis acceleration (Ax in the legend) is plotted over time, with time discretized based on the 10 ms acquisition periodicity of the motion sensor. The yellow vertical bar indicates the instant of the anticipated \gls{ai}-based prediction, while the gray label marks the actual moment of the spill. When a risk is detected, the \gls{ai} system utilizes the \gls{5g} connection to activate a custom-built gripper to secure the liquid (represented in our tests by a water bottle), thereby ensuring operator safety and maintaining operational efficiency. In c), the left image shows the result when the \gls{poc} is inactive (i.e., the water bottle falls due to a sudden obstruction), while the right photo demonstrates how the proposed \gls{iiot} solution successfully prevents this outcome by accurately predicting the spill and activating the gripper in advance.}
 }
	\label{fig:poc_mir}
\end{figure*}

\noindent {\color{black} \textbf{Security and Safety}. The last point we want to highlight is the need for robust encryption algorithms and authentication mechanisms in all \glspl{ad} to ensure data integrity and confidentiality within industrial wireless networks, safeguarding sensitive information \cite{AngValMon:J22}. Equally critical, especially in industrial environments involving human-machine interaction, is functional safety, which focuses on preventing hazardous failures by ensuring reliable operation of safety-critical systems, even under fault conditions (see Fig.~\ref{fig:poc_mir}). From a communication standpoint, achieving functional safety requires dedicated safety communication protocols capable of verifying the integrity, authenticity, and timeliness of transmitted data. Addressing these challenges necessitates integrating safety communication protocols with encryption and authentication techniques to meet stringent \gls{e2el} requirements.}

\textit{Open challenges:} In addition to the previously mentioned need for new encryption and authentication techniques, the problem of designing fast and accurate intrusion detection systems, capable of monitoring large portions of the spectrum, and identifying and mitigating potential threats (i.e., jammers) or unauthorized access attempts is still open.
Finally, as quantum computing advances, the threat to traditional cryptographic methods grows. Preparing for post-quantum security involves deploying encryption algorithms resistant to quantum attacks, ensuring long-term data protection.

\red{As an experimental demonstration of some of these research directions, Figure \ref{fig:poc_mir} shows the results of one of our \gls{poc} that leverages the interplay between \gls{ai} and \gls{5g} \gls{nr} for enhancing safety in a real-world \gls{uc} belonging to the \gls{ad}4 family.}

\subsection{Modeling Guidelines}


In the following, we present modeling guidelines for industrial environments that adhere to the \gls{iiot} \red{characterization given by this paper}. This approach holds significant importance in establishing a \red{cohesive} framework for scientific endeavors aimed at addressing the research gaps previously highlighted. Indeed, the applicability of existing literature contributions to \gls{iiot} environments strongly depends on the level of realism of their underlying assumptions and characterizations. Moreover, the comparison between different solutions is possible only when they rely on \red{common} models.

Specifically, in our vision, the performance evaluation of networking protocols, algorithms, mathematical models, and other solutions is highly influenced by several key factors that characterize the industrial environment under consideration. Consequently, we will thereby describe each factor in detail, by providing indications on how to model them and examples of how to apply these rules for the considered \gls{iiot} \glspl{ad}. 

\paragraph{Number of devices} 
The number of devices strictly depends on the type of \gls{ad}. In general terms, it can be either fixed or variable. In the former case, the number is typically decided based on the number of industrial assets that should be monitored or controlled via wireless devices. However, this complete knowledge of the environment is difficult to achieve in reality as it requires indications from industrial partners. Hence, the most common approach is to consider an average number of devices per square meter, such that the total number of devices can be obtained according to a Poisson distribution.

\paragraph{Service area}

The service area is the overall physical space where the \gls{ad} is applied. In the case of \glspl{ad} referring to a single building (e.g., \gls{ad}1), the typical building size ranges from 100 m$^2$ to 1 km$^2$, whereas the building height is from 5 m to 25 m \cite{3gpp22804}. Conversely, when considering entire factories (e.g., \gls{ad}4), the service area can be much larger, i.e., up to 100 km$^2$. These reference numbers are drawn from both existing literature, 
and real industry plants\footnote{See, e.g., those affiliated with the BIREX Competence Center for the Industry 4.0, which can be accessed at: \url{https://bi-rex.it/en/partners/}.}.



\paragraph{Device distribution}

An important key factor when characterizing the industrial \glspl{ad} refers to the spatial models, that is, indicating how the devices can be located in the service area. Similar to the considerations made for the number of devices, the most common spatial distribution models are deterministic or random. In the former case, additional knowledge of the industrial environment is needed (e.g., the position of the industrial assets), whereas, in the latter case, devices are typically distributed uniformly in either the entire service area or spatially limited in specific regions of the factory and/or building (e.g., when considering sensors mounted on a given machine). Notice that joint considering of both a random number and spatial distribution of devices entails contemplating a \gls{ppp}. 
It is worth highlighting that \glspl{bs} should rather be situated deterministically to provide a sufficient level of coverage to all devices.

\paragraph{Mobility}

Information regarding the potential speed and movement types of the different industrial entities enhances the ability to characterize specific aspects, such as the radio channel, which in turn impacts network performance. In industrial environments, there are mainly three types of mobility. Devices can be (i) stationary (e.g., motion controllers or \glspl{MCP} for \gls{ad}1 and \gls{ad}3), (ii) moving along pre-defined routes (e.g, \glspl{agv} for \gls{ad}4) or (iii) moving in a correlated way (e.g., \glspl{uav} for \gls{ad}4). Besides the straightforward characterization of case (i), the most common mobility model for case (ii) is the random walk stochastic process, 
where devices move from their current location by randomly choosing direction and speed. Typically, the speed is chosen randomly from 10 to 70 km/h\cite{3gpp22104}. Regarding case (iii), the most common correlated mobility models fall into two main categories: i) reference-based models, such as the Reference Point Group Mobility (RPGM) model, that determine the positions of devices with respect to the trajectory of a common reference point; ii) behavioral models, where the correlation between mobility patterns emerges by the application of a common set of rules to all devices. Recently, models that merge the simplicity of reference-based models with the accuracy and flexibility of rule-based models were also proposed \cite{Mo3}.

\paragraph{Traffic models}

The modelling of traffic types is another fundamental aspect to be considered. It describes how data are generated by devices. Typically, data generation by a traffic source is modeled as a stochastic process with two main random variables, (i) size of the generated data (i.e., \gls{ms}), and (ii) time interval between two subsequent data generation instants (i.e., \gls{ti}). The values of \gls{ms} and \gls{ti} can be either fixed according to specific \glspl{kpi} (see Table \ref{table_requirements}), or variable. In the latter case, the distribution of \gls{ms} and \gls{ti} is typically obtained by collecting traffic traces and analyzing the behavior of real industrial applications. Alternatively, common stochastic models such as Poisson, Negative Exponential, and Generalized Beta distributions are commonly employed. Intriguingly, spatio-temporal Markov Decision Processes also serve as a viable representation of the correlations among data generated by proximate devices (e.g., sensors associated with the same industrial machine) \cite{cavallero2023}.  

\paragraph{Channel models}

Industrial environments possess distinctive characteristics that render wireless propagation inherently unique and distinguishable from other radio channels. Typically, industrial premises exhibit larger dimensions and/or greater height compared to office or residential buildings. 
Furthermore, the size, density, and spatial distribution of industrial equipment can vary significantly from one case to another but, on average, constitute a highly dense scatterer environment. In light of these peculiarities, it becomes imperative to tailor channel models to suit industrial scenarios adequately. For instance, the \gls{3gpp} has proposed a path loss model specifically designed for indoor factories\footnote{It is noteworthy that the term \virg{factory} in \gls{3gpp} corresponds to the term \virg{building} used in this article.}\cite{3gpp38901}. However, more detailed wideband approaches can be obtained by leveraging (i) ray-tracing tools, as propagation in factories is likely to be more site-specific than propagation in usual residential or office environments, 
(ii) channel measurements, as a proper calibration method for ray-tracing tools, or (iii) the utilization of \gls{ai}. Indeed, the latter is recently gaining attention to discern the intricate relationship between common propagation markers and the key properties of industrial environments \cite{seretis2021overview}.

{\color{black} We would also like to remark that experimental validation of models is inherently challenging in complex industrial systems due to limitations in current measurement techniques and data acquisition methods, which often struggle to capture the multidimensional interactions and dynamic behaviors necessary for comprehensive validation.}

\section{Conclusions}
\label{sec:6_Conclusions}

In this article, we presented a \red{comprehensive characterization} of the \gls{iiot} framework, a key enabler of the emerging fourth industrial revolution. Our focus included \red{clarifying the diverse and sometimes conflicting} contributions from entities like \gls{3gpp} and \gls{5gacia} in terms of nomenclature, \glspl{ad}, \glspl{kpi}, and traffic types. After a careful literature analysis, we also \red{outline} the main \gls{iiot} requirements by harmonizing heterogeneous terminologies. Most importantly, we offered an overview of the main research topics and challenges within the \gls{iiot} paradigm according to the \red{evolution of mobile radio networks}, highlighting the potential for innovative industrial applications. Finally, we proposed modeling guidelines as a foundational framework for future research endeavors, addressing the identified research directions towards next-generation \gls{iiot}.




\bibliographystyle{IEEEtran}
\bibliography{biblio.bib}

\section*{Biographies}

\textbf{Giampaolo Cuozzo} [M'23] (giampaolo.cuozzo@wilab.cnit.it) received from the University of Bologna a B.S with honors in electronics and telecommunications engineering in 2017, an M.S with honors in telecommunications engineering in 2019, and a Ph.D. degree in electronics, telecommunications, and information technologies engineering (ET-IT) in 2022. He is currently Head of Research at the National Laboratory of Wireless Communications (WiLab) of CNIT (the National, Inter-University Consortium for Telecommunications). His research activity is focused on the study, development, and validation of 6G networks for the Industrial Internet of Things.

\textbf{Enrico Testi} [M'22] (enrico.testi@unibo.it) received the M.S. degree in electronics and telecommunications engineering for energy (magna cum laude) and the Ph.D. degree in Electronics, Telecommunications, and Information Technologies engineering from the University of Bologna, Italy, in 2018 and 2022, respectively. He is currently junior assistant professor at the Department of Electrical, Electronic, and Information Engineering \virg{Guglielmo Marconi} of the University of Bologna. His research interests include artificial intelligence techniques for next-generation wireless networks, massive MIMO, and satellite IoT.

\textbf{Salvatore Riolo} [M'22] (salvatore.riolo@unict.it) received an M.S (with honoris) in telecommunications engineering, and a Ph.D. degree in systems, energy, computer and telecommunications engineering from the University of Catania, Italy, in 2017 and 2021, respectively. He has been a junior assistant professor of telecommunications with the Department of Electrical, Electronics and Computer Engineering (DIEEI), University of Catania, since January 2022. His scientific interests include radio resource management for B5G/6G networks, massive machine-type communications, and the convergence of artificial intelligence with radio access networks.

\textbf{Luciano Miuccio} [M'22] (luciano.miuccio@unict.it) received an M.S (with honoris) in telecommunications engineering, and a Ph.D. degree in systems, energy, computer and telecommunications engineering (with the additional label of “Doctor Europaeus”) from the University of Catania in 2018, and 2022, respectively. He is currently a junior assistant professor at the Department of Electrical, Electronics and Computer Engineering of the University of Catania. His scientific interests include green networking, NOMA techniques, and AI for B5G/6G radio resource management.

\textbf{Gianluca Cena} [SM'09] (gianluca.cena@ieiit.cnr.it) received an M.S. degree in electronic engineering and a Ph.D. degree in information and system engineering from the Politecnico di Torino, Italy, in 1991 and 1996, respectively. Since 2005 he has been a Director of Research with the National Research Council of Italy (CNR-IEIIT). His research interests include wired and wireless industrial communication systems, real-time protocols, and automotive networks. Since 2009 he has been an Associate Editor of the IEEE Trans. Industr. Inform.

\textbf{Gianni Pasolini} [M'00] (gianni.pasolini@unibo.it) holds an M.Sc. in Telecommunications Engineering and a Ph.D. in Electronic Engineering and Computer Science from the University of Bologna, Italy. Currently, he is an Associate Professor in the same University, where he has been teaching telecommunications courses since 2003. His research focuses on wireless communication systems, the Internet of Things, digital signal processing, and THz communications. He is a founding member of the \virg{National Laboratory of Wireless Communications - Wilab} at the National Inter-University Consortium for Telecommunications (CNIT). He serves as an Associate Editor for IEEE OJ-COM.

\textbf{Luca De Nardis} [M'98] (luca.denardis@uniroma1.it) is an Associate Professor with the Department of Information Engineering, Electronics and Telecommunications, Sapienza University of Rome. In 2007, he was a Postdoctoral Fellow with the University of California at Berkeley, Berkeley. He authored or coauthored over 120 international peer-reviewed publications. His research interests include cognitive communications, medium access control, routing protocols, and wireless positioning systems.

\textbf{Daniela Panno} [M’19] (daniela.panno@unict.it) received an M.S (Hons.) in electrical engineering from the University of Catania, Italy, in 1989, and a Ph.D. degree in electronic engineering and computer science engineering from the University of Palermo, Italy, in 1993. Since 1998, she has been an Associate Professor of telecommunications at the University of Catania. Her current research interests include massive IoT scenarios, green networking, and AI-based radio resource management for B5G/6G networks.

\textbf{Marco Chiani} [F'11] (marco.chiani@unibo.it) is currently a Full Professor of telecommunications at the University of Bologna. His research interests include information theory, wireless systems, statistical signal processing, and quantum information. Since 2003, he has been a Frequent Visitor with the Massachusetts Institute of Technology (MIT), Cambridge, where he holds a Research Affiliate appointment. He received the 2011 IEEE Communications Society Leonard G. Abraham Prize in the Field of Communications Systems, the 2012 IEEE Communications Society Fred W. Ellersick Prize, and the 2012 IEEE Communications Society Stephen O. Rice Prize in the Field of Communications Theory.

\textbf{Maria-Gabriella Di Benedetto} [F'16] (mariagabriella.dibenedetto@uniroma1.it) is a Full Professor in telecommunications at Sapienza University of Rome. She is currently a Research Affiliate with the Research Laboratory of Electronics (RLE), MIT. She is a fellow of the Radcliffe Institute for Advanced Study, Harvard University, Cambridge, MA, USA. In 1994, she received the Mac Kay Professorship Award from the University of California at Berkeley. Her research interests include wireless communication systems, impulse radio communications, and speech. 

\textbf{Enrico Buracchini} (enrico.buracchini@telecomitalia.it) is an electronic engineer who joined Telecom Italia in 1994 to focus on radio innovation across various locations of the company, including Italy, Austria, Greece, and Spain. Today, as a senior project manager, he coordinates the activities related to the evolution of 5G, having managed the project and several trials, including one in San Marino. Furthermore, as an expert speaker, he collaborates with TIM's Business Directorate in meetings with clients interested in 5G solutions. Currently, he represents TIM in international standardization groups, including 3GPP RAN1 and ITU R 5D.

\textbf{Roberto Verdone} [SM'18] (roberto.verdone@unibo.it) is a full professor at the University of Bologna, since 2001. He is the Director of WiLab, the Italian Laboratory of Wireless Communications of CNIT. He is also co-director of the Joint Innovation Center on "Intelligent IoT for 6G" with Huawei. His main research interests are on the evolution from 5G to 6G, and the Internet of Things. He published 200 scientific papers and a few books on various aspects of wireless communications.

\end{document}